\documentclass{ws-p8-50x6-00}

\begin{document}

\title{Selection Effects in high-redshift 
Submillimeter Surveys and pointed observations} 

\author{Andrew W. Blain}

\address{Institute of Astronomy, Madingley Road, Cambridge, CB3 0HA, UK
\\E-mail: awb@ast.cam.ac.uk}


\maketitle

\abstracts{
The results of the first generation of submillimeter (submm)-wave surveys 
have been published. The opening of this new window on the distant Universe 
has added considerably to our understanding of the galaxy formation process, by  
revealing a numerous population of very luminous distant galaxies. Most would 
have been very difficult to identify using other methods. The potential importance
of selection effects, especially those connected with the spectral energy distributions 
(SEDs) of the detected galaxies, for the interpretation of the results are highlighted. 
} 

\section{Introduction}

From the {\it IRAS} survey, it was clear that the absorption and re-emission of 
starlight by interstellar dust in the Milky Way, and of both starlight and accretion energy 
from AGN by dust in external galaxies, is a very important process. The thermal emission 
spectrum of dust heated by this radiation, which is observed to peak at a wavelength 
of order 100\,$\mu$m, is redshifted into the submm waveband with a very strong 
$K$-correction, making high-redshift submm galaxies unusually easy to detect as 
compared with their low-redshift counterparts. Between redshifts of about 0.5 and 10 
the $K$-correction almost balances the cosmological dimming of a source with a fixed 
template SED, and so the flux density received from a galaxy is approximately
constant.\cite{Hyfest} Although the submm waveband is thus a very attractive window 
for cosmology,\cite{BL93} it is technically very challenging to image submm radiation at 
the faint sensitivity levels -- several mJy at 850\,$\mu$m -- required to detect even an 
ultraluminous galaxy, defined as possessing a far-infrared(IR) luminosity in excess of 
$10^{12}$\,L$_\odot$, at any redshift $z \ge 0.5$. 

Systematic submm-wave cosmology has only been possible since the 
commissioning of the SCUBA camera at the James Clerk Maxwell Telescope in 
1997.\cite{Holland} A handful of known high-redshift galaxies and 
AGN were detected earlier using single-pixel detectors,\cite{10214,4C,BR1202} but 
blank-field surveys were impossible, due to a combination of relatively low 
sensitivity and a very small field of view. SCUBA brought 37 and 91 detectors operating 
in the atmospheric windows at 850 and 450\,$\mu$m respectively, providing a 
2.5-arcmin field of view, and has been used to good effect to make a range 
of surveys of the high-redshift 
Universe.\cite{SIB,Hughes,Barger98,Eales,BCS,PvdW,Chapman} Recently, the MAMBO 
camera at the IRAM 30-m telescope has also produced deep survey images 
at 1.25\,mm.\cite{Bertoldi} The existing BIMA, IRAM, Nobeyama and OVRO interferometer 
arrays are very valuable for making sensitive, high-resolution mm-wave observations, 
but because of their small fields of view, they are not practical survey instruments. Prior 
to the debut of SCUBA, the one-dimensional multi-channel 350-$\mu$m SHARC camera at 
the Caltech Submm Observatory (CSO) was used to limit the counts of faint submm 
galaxies, and the relatively large primary beam of the BIMA array at 2.8\,mm was exploited 
to impose the first limit to the mm-wave counts in a mosaicked image of the Hubble 
Deep Field.\cite{WW} The detection of a very significant intensity of 
submm/far-IR background radiation from the {\it COBE} FIRAS and DIRBE datasets 
was achieved in parallel to the first SCUBA surveys.\cite{Puget,SFD,Hauser,FSD} 

The prospects for further instrumental developments are excellent. The lithographic 
manufacture of large arrays of bolometers is now routinely demonstrated.\cite{Bock} 
The BOLOCAM detector array\cite{Glenn} that uses one type of this technology has 
recently undergone its first engineering tests at the CSO. An alternative technology is 
exploited in the forthcoming SHARC-II\cite{Dowell} camera for the CSO. Bolometers that 
exploit superconducting and quantum interference devices rather than simple 
thermistors are also being developed. These promise increased stability and reduced 
response time, and would lead to a crucial increase in the degree of multiplexing 
possible in their readouts, and so to much larger arrays. The future SCUBA-II\cite{SCUBA2} 
camera for the JCMT and large-format bolometer cameras for the 50-m LMT\cite{Schloerb}
are expected to exploit such devices. The SMA\cite{Wilner_here} interferometer array on 
Mauna Kea will provide the first sensitive, fully-$uv$-sampled interferometric 
images in the submm band, and in the future the 64$\times$12\,m ALMA 
array\cite{Wootten} will provide extremely sensitive, high-resolution observations.  

Wide-band mm/submm-wave spectrometers\cite{WASP} are also being 
developed, using both arrays of heterodyne detectors\cite{Erickson} and dispersive 
techniques.\cite{BFBS} These instruments will offer the potential for the direct 
determination of redshifts for mm/submm-selected galaxies from mm-wave
observations of CO lines. There is a natural symbiotic relationship between new 
panoramic bolometer cameras\cite{Glenn,SCUBA2} and these spectrographs, 
which together will be capable of both detecting and obtaining redshifts for large 
samples of high-redshift dusty galaxies without recourse to either optical, near-IR, or 
even radio telescopes.

The results and consequences of the first SCUBA surveys have been discussed 
extensively elsewhere.\cite{BSIK,Hughes,Eales,Lilly,BJSLKI,BCS,Lilly_here,Smail_UMass} 
Here the potential selection effects in this new waveband\cite{Eales,Lilly,Hyfest} are 
described in the context of our knowledge of the SEDs of the detected objects. 

\section{Submm-wave selection effects} 

\subsection{Redshifts of submm-selected galaxies} 

In mm/submm-wave surveys there is a unique bias in favor of the 
detection of high-redshift galaxies at the expense of their low-redshift 
counterparts. The flux density of a galaxy {\em with a fixed SED} is expected
to be approximately constant over a wide range of redshifts from about 0.5 to 10. The 
increase in the volume element out to $z \simeq 2$ conspires to bias the selection 
function to greater redshifts; a demonstration of this effect in currently 
popular world models as a function of SED is presented elsewhere.\cite{Hyfest} 

There is broad agreement\cite{Lilly_here,Smail_UMass} that the objects detected in 
SCUBA surveys are at redshifts $z \simeq 3 \pm 2$. All three SCUBA galaxies with 
redshifts confirmed using CO spectroscopy\cite{Frayer1,Frayer2,Soucail} are at $z > 1$. 
Others for which deep multiwaveband data is 
available\cite{Hughes,Downes,ERO,Eales,Gear} have no potential low-redshift 
counterparts. Four of the first galaxies from the CUDSS survey\cite{Eales-II,Lilly} were 
identified with galaxies at $z < 1$; however, more extensive results\cite{Eales,Lilly_here} 
show a smaller fraction of potential low-redshift identifications. Note that two initially 
plausible low-redshift identifications of galaxies in the SCUBA Lens
Survey\cite{SIB,SIBK} were subsequently revised upwards in the light of additional 
data.\cite{ERO} New results from this meeting\cite{PvdW,Chap} are consistent 
with a distant redshift distribution. 

Despite their potential for detecting high-redshift galaxies, submm surveys detect 
galaxies at restframe wavelengths considerably longwards of the peak of their SEDs. It 
is thus possible that the form of this SED can affect the interpretation of
the results of SCUBA surveys.\cite{Eales,Hyfest} The main effect 
would be to overestimate the bolometric luminosity associated with a
submm-selected galaxy if its dust temperature was overestimated. 

\subsection{SEDs of submm-selected galaxies}

Potential selection effects in submm-wave surveys are controlled by the SEDs of 
the detected galaxies. The SEDs of individual dusty 
galaxies certainly vary from galaxy to galaxy, and could vary systematically with 
luminosity and redshift. Here and in previous work connected with the SCUBA Lens 
Survey,\cite{BSIK,BJSLKI,Hyfest} three parameters are used to describe the SED; a dust 
temperature $T$ in a standard Planck function $B_\nu$, an index $\beta$ in a dust 
emissivity function $\nu^\beta$ and a spectral index $\alpha$ to describe the mid-IR SED,
$f_\nu \propto \nu^\alpha$. 

Dust SEDs are steeper than blackbody spectra in the Rayleigh--Jeans (RJ) regime. If 
the SED is represented by the function $\nu^\beta B_\nu$, then the RJ spectral index 
is $-2-\beta$. Some authors\cite{Benford,LIH} represent the SED by a greybody of the 
form $(1-\exp{-K \nu^\beta})B_\nu$, explicitly taking account of the increase in optical 
depth with increasing frequency. In this case, the RJ spectral index is also $-2-\beta$, 
but the SEDs differ near their peaks. Temperatures fitted using a greybody SED are 
typically about 20\% greater than those derived from the $\nu^\beta B_\nu$ SED. 

The mid-IR $\alpha$ term is introduced because dust SEDs do not fall exponentially like 
blackbodies in the Wien regime, due to the contribution from additional hot 
dust components in the interstellar medium. At frequencies above which the gradient 
of $\nu^\beta B_\nu$ is steeper than that of the power law $\nu^\alpha$, the SED is 
represented by this power law, thus incorporating the observed mid-IR properties of 
galaxy SEDs; see Fig.\,1. The radio emission associated with dusty galaxies is represented 
by an additional power-law function, normalized to fit the low-redshift far-IR--radio 
correlation.\cite{Condon}

These three simple parameters can provide an adequate and appropriate description of 
the SEDs of distant dusty galaxies from the mm to mid-IR waveband,\cite{diffmag} which  
are constrained only by a handful of broad-band photometric measurements with little 
or no spatial resolution. When sub-arcsecond spatially-resolved images from ALMA are 
available, it will be important to investigate more details of the mm to near-IR SEDs, 
using radiative transfer models, including full details of the complex geometry, the 
different physical components and conditions in a merging, star-forming galaxy, and of a 
potential point-like nuclear heat source.\cite{E&RR,Guiderdoni} However, few of the 
parameters required by such models can be constrained at present, and so we prefer to 
use the simple three-parameter SED. Based on fits to the form of evolution of 
low-redshift 60-$\mu$m {\it IRAS} galaxies and more distant galaxies selected at 
175\,$\mu$m, described elsewhere,\cite{BSIK} values of $T=38$\,K, $\beta=1.5$ and 
$\alpha=-1.7$ were chosen; the associated SED is shown by the dashed line in Fig.\,1.  

Information about the SEDs of the high-redshift luminous galaxies detected in 
submm surveys can be obtained by several different routes. 

First, submm-wave observations of a representative sample of galaxies from the 
{\it IRAS} catalog can be made.\cite{Dunne} However, being at low redshifts ($z \le 0.1$), 
the properties of these galaxies have rather little overlap with those of the more 
distant, typically much more luminous galaxies found in deep submm surveys. The 
results indicate that there is a weak trend for the dust temperature to increase with 
increasing luminosity, and that $T=36 \pm 5$\,K and $\beta = 1.3 \pm 0.2$ are reasonable 
typical values; see the dotted SED shown in Fig.\,1. Independent results for the typical 
SEDs of {\it IRAS} galaxies, which yields a similar generic spectrum, are shown by the 
solid line in Fig.\,1.\cite{Guiderdoni-MN}

Secondly, color information for high-redshift submm-selected dusty galaxies can be 
obtained from multiband submm/mm/IR observations and combined with
independent redshift information to derive $T$ and $\beta$.\cite{I+7,I+7-II} The few 
temperatures available are in the range 40-50\,K, consistent with the SEDs described 
above. The relevant data are plotted in Fig.\,1.\cite{Hyfest,HyLIG} Note that because dust 
SEDs are thermal, it is impossible to distinguish {\it a priori} between the effects of an 
increase in temperature or a reduction in redshift.

Thirdly, the color technique can be used to measure SEDs for high-redshift galaxies 
with known redshifts, but a less uniform selection criterion, for example the most 
luminous {\it IRAS} galaxies, high-redshift radio galaxies, optically-selected AGN and 
gravitational lenses.\cite{Hyfest,HyLIG} The ranges of luminosity and 
redshift that are appropriate to the SCUBA galaxies are sampled using this approach; 
however, the surveyed objects are diverse, and perhaps extreme and unrepresentative. 
For example, the dust temperature inferred from observations of the most luminous 
lensed quasar\cite{Lewis,diffmag} is about three times greater than the typical 
temperature inferred from {\it IRAS}, {\it ISO} and SCUBA data. 

In general, dust temperatures in high-redshift AGN seem to be greater than in 
{\it IRAS} and SCUBA galaxies. This is probably due both to the selection effect against 
detecting hot objects in SCUBA surveys, and to higher intrinsic temperatures in the most 
luminous objects. 

\subsection{Pointed observations} 

There has been much concerted effort to detect statistical samples of known 
high-redshift galaxies, such as the Lyman-break galaxies,\cite{Chap} classical radio 
galaxies,\cite{Archibald} quasars\cite{Benford,Isaak} and gravitational lenses using 
mm/submm instruments. Before SCUBA, this technique provided the only opportunity to 
study the high-redshift mm/submm Universe.\cite{10214,4C,BR1202} Today, it provides 
a direct opportunity to connect the results obtained from deep submm surveys to more 
developed fields of observational cosmology carried out in other wavebands.

The selection functions at work in these studies favor the detection of the highest 
redshift members of flux-limited samples of objects selected in another waveband, 
because of the familiar mm/submm $K$-correction. The ratio between the submm flux 
density of a galaxy and that measured in any other waveband, with the possible 
exception of soft X-ray observations of highly-enshrouded gas-rich AGN, is expected to 
be a strongly increasing function of redshift. For example, the submm--radio/optical 
flux density ratio of a galaxy with a reasonable radio or optical SED, 
$f_\nu \propto \nu^{-1}$, and a submm SED, $f_\nu \propto \nu^{3.5}$, should increase 
with redshift $z$ as $(1+z)^\gamma$, where ${\gamma \simeq 4.5}$. The SCUBA detection 
of a $z=5.8$ quasar discovered in the Sloan Digital Sky Survey reported at this meeting, 
despite the typical lack of detections of $z \simeq 3$ quasars,\cite{Isaak} provides a 
possible example. It is exciting that the growing number of very high-redshift objects 
detected in other wavebands could be relatively easy to detect in pointed mm/submm 
observations, using existing telescopes like the JCMT and CSO, forthcoming telescopes 
like the SMA and LMT, and ultimately at great sensitivity and resolution using ALMA. The 
intergalactic medium is always transparent to mm/submm radiation, and so this type of 
observation could be very important for tracing the process of re-ionization in the years 
ahead. 

It should be possible to study the intrinsic evolution of the dust emission properties of a 
well-defined sample of galaxies by making mm/submm observations of a
sub-sample that have been selected carefully elsewhere to have a fixed luminosity as a 
function of redshift.\cite{Archibald} For example, while the fraction of 
powerful radio galaxies drawn from flux-limited samples that are detected by SCUBA is 
observed to increase with redshift, consistent with the selection effect above, a 
sub-sample with matched radio luminosities still displays a statistically significant 
increase in dust luminosity with redshift, indicating that powerful radio galaxies are 
systematically more luminous in the submm than the radio waveband at higher 
redshifts.\cite{Archibald}
 
\section{Summary: future observations of very early galaxies} 

Selection effects in deep mm/submm-wave surveys generally favor the detection of 
the most distant objects. For a fixed bolometric luminosity, dusty galaxies with colder 
dust temperatures are more likely to be detected. The detection of galaxies at the 
highest redshifts in the mm/submm waveband requires only that they are luminous 
and contain dust. Because absorption along the line of sight is not significant in the 
mm/submm waveband, dust generated and heated in the very first galaxies, prior to the 
epoch of re-ionization, could potentially be detected in the submm waveband. 

\begin{figure}[t]
\begin{center}
\epsfig{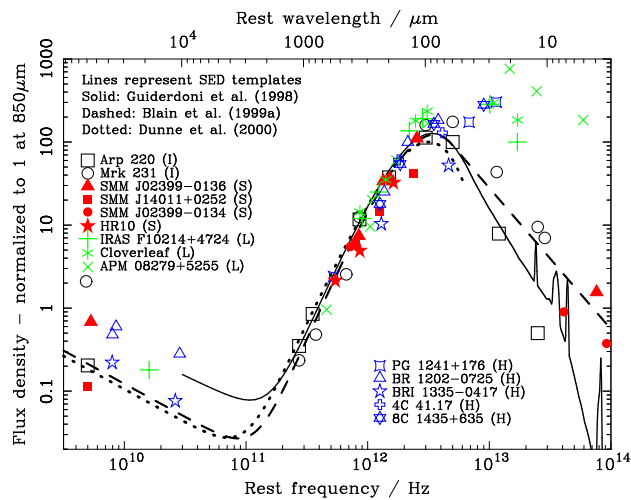}
\end{center}
\caption{SEDs of a wide range of dusty galaxies with known redshifts, 
including known lensed sources (L), high-redshift AGN (H), low-redshift 
{\it IRAS} galaxies (I) and submm-selected galaxies (S). Three spectral 
templates are also shown. Only those objects with redshifts and both mid-IR 
and submm data are plotted. Both the re-normalized data and 
the templates are very similar at wavelengths longer than 60\,$\mu$m. 
\label{fig:radish}}
\end{figure}

\section*{Acknowledgments}
I thank the Raymond and Beverly Sackler Foundation for support at the IoA, 
UMass/INAOE for support at the meeting, the CfA for organizing `The First 
Generation of Cosmic Structures', for which some of this material was prepared, 
and Jan and Rick Caldwell for hospitality prior to both meetings.

\end{document}